\documentclass[
 amsmath,amssymb,
 aps,
 prfluids
]{revtex4-2}

\usepackage{xspace}
\usepackage{graphicx}
\usepackage{dcolumn}
\usepackage{bm}
\usepackage{natbib}
\usepackage{color}
\usepackage{float}
\usepackage{xcolor}
\usepackage{ulem}

\newcommand{\music}{\texttt{MUSIC}\xspace}
\newcommand{\Msun}{M_\odot}

\begin{document}

\preprint{APS/123-QED}

\title{Convection in the interior of Cepheid variable stars}

\author{Maxime Stuck}
 \email{stuck2@llnl.gov}
\author{Jane Pratt}%
 \email{pratt34@llnl.gov}
\affiliation{%
 Lawrence Livermore National Laboratory, 7000 East Ave, Livermore, CA 94550, USA}%


\date{\today}
\begin{abstract}
This paper is associated with a poster winner of a 2025 American Physical Society's Division of Fluid Dynamics (DFD) Gallery of Fluid Motion Award for work presented at the DFD Gallery of Fluid Motion. The original poster is available online at the Gallery of Fluid Motion, \url{https://doi.org/10.1103/APS.DFD.2025.GFM.P005}.
\end{abstract}

\maketitle


In stars, the description of shell convection presents scientific challenges not encountered in either core convection or envelope convection.  Convection in each of these settings has the potential to significantly overshoot the boundaries of the convective instability, penetrating deep into stable layers.  However, in shell convection, the same convection roll can overshoot both the upper and lower boundaries of the shell with each rotation.   In this work, we describe the evidence from extensive visualizations of shell convection, and compare that to core overshooting and envelope overshooting.

Convective shells are common features of Cepheid variable stars, as well as other stars on the asymptotic giant branch (AGB).   Cepheids undergo periodic variations of their radii and luminosity with timescales ranging from a few days to a few weeks. The period-luminosity relation \cite{leavitt1912periods} enables observational data of Cepheids to provide an indirect distance measurement, so that these bright variable stars are important across astronomy and cosmology.
The theoretical description of Cepheids has documented shortcomings \citep{guzik2023challenges}, evident through the Cepheid mass discrepancy problem. A systematic disagreement of $10\%$ to $20\%$ is found when comparing Cepheid masses inferred from stellar evolution models with those obtained from astronomical observations or dynamical constraints \citep[e.g.,][]{bono2006stellar, cassisi2011classical}.  In trying to match the observed luminosities of Cepheids, one-dimensional evolutionary models produce stars that are more massive than observations indicate. This suggests that the internal transport of energy may not be modeled with sufficient convective efficiency \citep[e.g.,][]{neilson2011cepheid}. Among the mechanisms proposed to explain this discrepancy is convective overshooting. Because convective motions have momentum, they can cross the theoretical boundaries of the convective instability, thereby mixing material and redistributing energy within adjacent stable layers. Because overshooting occurs in the deep interior of a star, it cannot be measured by observations; multi-dimensional hydrodynamic simulations can allow the detailed interaction between stable and unstable layers to be probed. When the radial resolution of the simulation is sufficient, the extent of an overshooting layer can be measured using the vertical kinetic energy flux as well as the thermal flux; for statistics gathered over long times, these different measurements converge \citep[for further discussion see][]{pratt2017extreme}.  The measured depth of the overshooting layer can then be used in one-dimensional stellar evolution models, impacting the Cepheid mass discrepancy problem.

Multidimensional hydrodynamic simulations provide a physically grounded approach, allowing the structure and dynamics of convective boundary mixing to emerge directly from the fluid equations. Here we examine Implicit Large Eddy Simulations \citep[ILES,][]{grinstein2007implicit} of the internal convective shell of Cepheids of different stellar masses that we have performed with the MUltidimensional Stellar Implicit Code (\music) \citep{pratt2016spherical, goffrey2017benchmarking}. The \music code solves inviscid compressible hydrodynamic equations, employing an implicit time-integration scheme that allows the efficient evolution of low Mach flows over long times.   While the full statistical analysis of these simulations is described in Stuck et al. \cite{stuck2025convective}, this work presents a study of the morphology of the flow through visualizations. We highlight the convective rolls and the extent of the convective overshooting layers in the surrounding radiative zones.  We compare the flow morphology to convective core convection and convective envelope convection.

\citet{stuck2025convective} studied stars of different mass within the Cepheid instability strip; the structures for these stars were produced by the open-source MESA code \citep[][]{paxton2010modules}. We adopt the common notation $\Msun$ to indicate the solar mass. For the $7 \Msun$ star, we have selected two different models, denoted (a) and (b), to allow for a comparison between similarly structured stars of the same mass at different ages. These 1D stellar models are used as initial conditions in our 2D hydrodynamic simulations. The six stellar models include a tiny convective core, an inner convectively unstable shell surrounded by two stable radiative zones, and a thin outer convective envelope.  Because our focus is overshooting around the convective shell, we truncate the simulation volume at a point just below the outer convective envelope, which removes the influence from overshooting below this external convection zone. A high radial resolution and a long time sequence of data are necessary to build converged statistics of this intermittent process. 2D simulations that assume axisymmetry generally produce higher convective velocities than three-dimensional (3D) simulations of stars.  The difference between 2D and 3D velocities is also dependent on the stellar structure.  Despite differences in the velocity, measured overshooting lengths are similar in 2D and 3D in \music simulations of the young sun \citep{pratt2020comparison,dethero2024shape}. Because they enable both higher radial resolution and the production of longer time series of data, we therefore have chosen to study 2D axisymmetric simulations in this work. 

The compressible fluid equations are solved in a spherical shell using spherical coordinates ($r$, $\theta$). We imposed periodicity on all physical quantities at the angular boundaries, as well as boundary conditions that maintain the original radial profiles of the 1D stellar evolution models in the radial direction. For a detailed examination of boundary conditions in \music, we refer to \citet{pratt2016spherical} and \citet{vlaykov2022impact}. All of the visualizations that we present here are produced during steady-state convection.

\begin{figure}[t]
\includegraphics[width=\columnwidth]{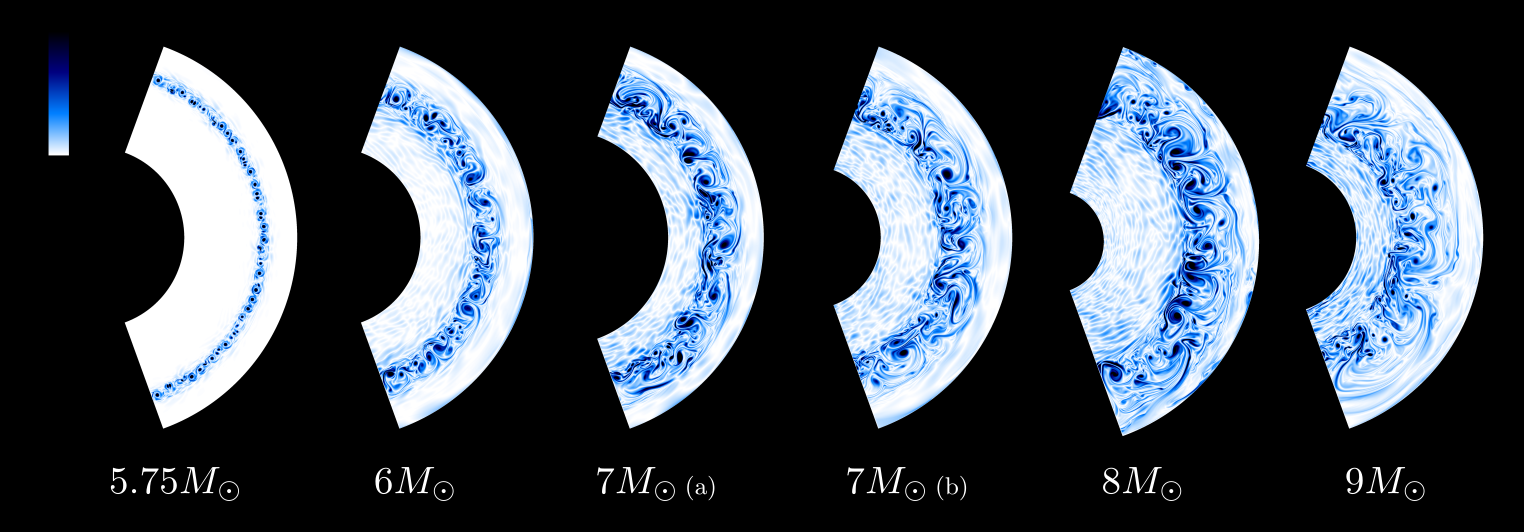}
\caption{\label{fig:vorticity} Visualization of vorticity magnitude for the six simulations of convective shells of realistic Cepheid variable stars at an arbitrary instant in time in steady-state convection. From left to right: $5.75 \Msun$, $6\Msun$, $7\Msun$ (a), $7\Msun$ (b), $8 \Msun$, and $9\Msun$. The zero point for the vorticity magnitude is white.}
\end{figure}
In our simulations of Cepheid models, an internal convective shell is clearly visible as a layer in radius where intense vorticity is present (see  FIG.~\ref{fig:vorticity}). 
In the surrounding layers that are stable to the convective instability, energy is primarily transported through radiation rather than the motion of the fluid. However, vorticity is not zero in these regions. The non-zero vorticity reflects the propagation of internal gravity waves excited by convective motions and overshooting plumes. For this sequence of stars, with increasing stellar mass, the boundaries of the convective shell become less sharply defined, indicating a growing influence of overshooting. In general, more massive stars tend to transport more energy through their interiors, leading to more vigorous convective motions and larger overshooting layers. 


Convection within these relatively thin shells is dominated by coherent rolls whose characteristic size is comparable to the shell thickness itself. In contrast to the more frequently studied situation of convective envelopes, where convection becomes truly multi-scale, these convective shells produce convection rolls of sizes that are tightly grouped around a single scale. This is similar to patterns commonly seen in laboratory experiments of Boussinesq convection. Smaller-scale structures only emerge near the convective boundaries, consistent with boundary-layer-like flow patterns.

\begin{figure}[t]
\includegraphics[width=\columnwidth]{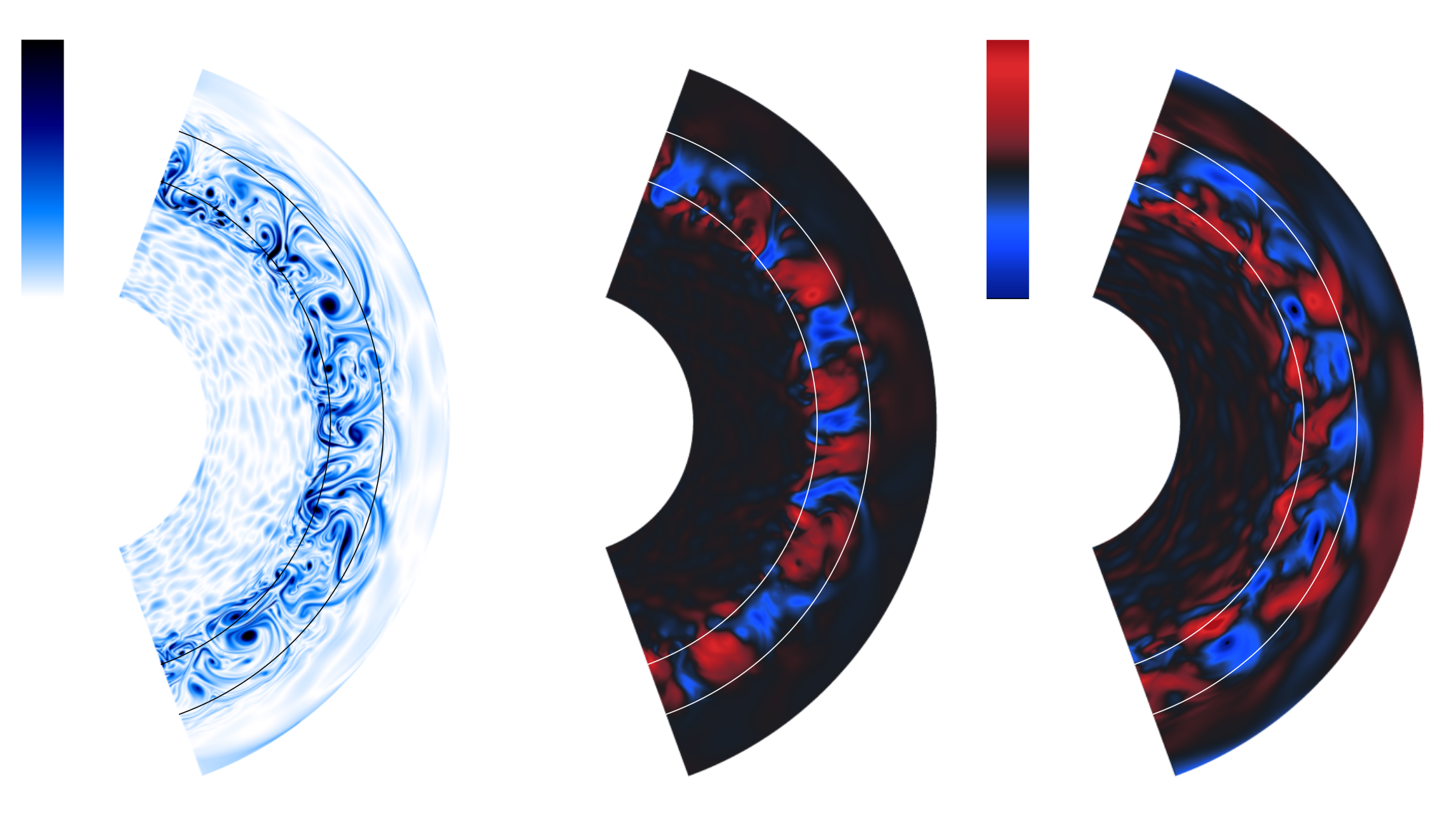}
\caption{\label{fig:visu_7b} Visualization of vorticity magnitude (left), radial velocity (center) and angular velocity (right) for the $7 \Msun$(b) Cepheid at an arbitrary instant in time in steady-state convection. The zero point for vorticity is in white in the left panel. In the center and right panels, outward flows are in red while inward flows are in blue; the zero point in velocity is black. The radial velocity ranges from $-1.0 \times 10^6$ cm/s to $8.5 \times 10^5$ cm/s at this instant in time, while the angular velocity ranges from $-9.8 \times 10^5$ cm/s to $1.1 \times 10^6$ cm/s. Solid lines indicate the position of the Schwarzschild boundaries for convective instability.}
\end{figure}
In a closer view of the $7 \Msun$ (b) simulation, convective flow structures can be seen to overshoot the boundaries of convective instability determined by the Schwarzschild criterion \citep{schwarzschild1958structure} (see FIG.~\ref{fig:visu_7b}).  The visualizations represent these convective boundaries by solid lines.  These boundaries are used in stellar structure and evolution models to determine the radial layer where stellar mixing length theory is applied.  Stellar mixing length theory assumes that convective motions stop abruptly at the convective boundaries, neglecting overshooting, and predicting a sharp transition between convective and radiative regions. This sharp transition is a well-known limitation of the common models used in 1D stellar structure and evolution. In contrast, in multidimensional simulations, convective boundaries are characterized by a continuous transition between convective and radiative regions.

The vorticity field shows that convection within the shell organizes into large convection rolls that extend beyond the formally unstable region into the adjacent stable layers. Overshooting is therefore a significant feature of these thin, shallow convective shells. The radial velocity field exhibits alternating inward and outward flows aligned with the vortical structures.  The angular velocity highlights the rotational character of the rolls. These visualizations illustrate that convection in thin stellar shells operates in a structured regime governed by a limited range of dominant scales. In the radiative layers, radial velocities rapidly decrease in magnitude, while thin bands of non-zero angular velocities remain visible. These correspond to internal gravity waves excited by convective motions. The wave patterns differ above and below the convective shell, a sign that their excitation by convective overshooting is asymmetric. The rounded features visible in FIG.~\ref{fig:visu_7b} correspond to large convection rolls protruding from each side of the convective shell. These contribute to the formation of large overshooting layers, in which the region just outside the convection zone is fully mixed.  More energetic convective structures can also overshoot further from the convective shell.

\begin{figure}[t]
\includegraphics[width=\columnwidth]{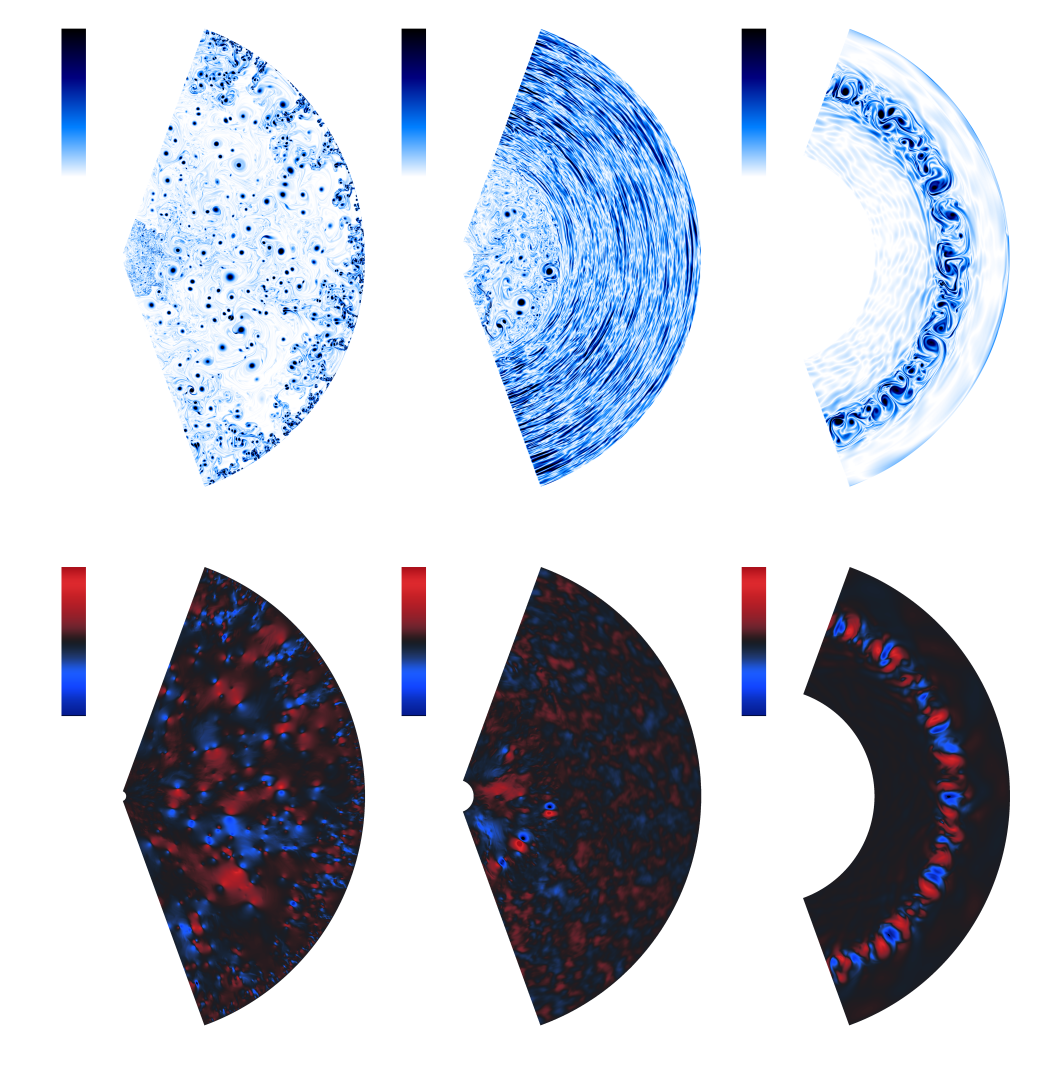}
\caption{\label{fig:comp_core_env} Visualization of vorticity magnitude (top), and radial velocity (bottom) for a convective envelope in a $4 \Msun$ pre-main-sequence star [left], a convective core of a $1.2 \Msun$ main-sequence star [center], and a convective shell in a $6 \Msun$ Cepheid variable star \citep[right,][]{stuck2025convective}.}
\end{figure}
To further illustrate the unique properties of stellar shell convection, we compare visualizations of our Cepheid simulations with a simulation of a deep convective envelope of a $4 \Msun$ pre-main sequence star, and a simulation of a convective core in a $1.2 \Msun$ main sequence star (see FIG.~\ref{fig:comp_core_env}).  All of these simulations were performed with \music using the same boundary conditions.  In convective cores and envelopes where the density can vary over orders of magnitude, convection rolls are excited with a broad range of interacting length and time scales. In this simulation of a convective core, the flow is dominated by small-scale, plume-like structures that fill the convectively unstable region. Approximate symmetry is observed between outward and inward flows. Highly asymmetric flows can be encountered in convective envelopes, characterized by fast, narrow inward flows and slow, wider outflows. In contrast, convection in thin shells is dominated by large, coherent rolls whose characteristic size visibly fills the convective shell's radial layer and dominates the flow. This narrow range of characteristic scales is consistent with the weak variation of the inflow widths across the convective region.  Quantitative analysis \citep{stuck2025convective} has shown that the inflow width varies from about $1\%$ to $3\%$ of the volume at a given radius, ultimately leading to convection rolls of nearly the same size in the convective shell.  This can be compared to the inflow widths in envelope convection reported by  \citet{dethero2024shape}, which varied between $1\%$ and $15\%$ of the volume for a red giant star near the first dredge-up point. The radial velocity fields (bottom row of FIG.~\ref{fig:comp_core_env}) emphasize these differences. Core and envelope convection display strongly multiscale patterns, while the thin shell regime shows alternating, spatially coherent bands associated with organized rolling motions. The narrow radial extent of the shell constrains the radial size of convective structures and therefore limits the range of dynamically accessible length scales, favoring coherent rolls whose size is set by the shell thickness. Shell convection thus represents a distinct hydrodynamic regime for stars.  We suggest that the differences in nonlinear gradients in density and temperature across the convectively unstable layer, as well as the extent of the convective layer, may contribute to these differences.

Our visualizations highlight the structured nature of convection, and the dynamical complexity of convective boundary mixing, in stellar convective shells.  This regime of stellar convection differs strongly from the convection found in deep envelopes or convective cores, where the length scales of flows are impacted by a strong stratification. By directly resolving these flows, multidimensional simulations provide essential insight into processes that remain fundamentally parameterized in one-dimensional stellar evolution models. The present simulations are 2D and assume axisymmetry.  3D simulations and whole-star simulations would be desirable as long as those simulations have a radial resolution equivalent to that of the simulations studied here. This work has focused purely on the hydrodynamics surrounding shell convection in realistic stratifications for Cepheid variable stars.  Future studies of these stars may focus on a multi-physics approach, including chemical burning, nucleosynthesis, rotation, and magnetic fields in the simulations. Understanding the interaction between convection and radial pulsations also remains an important direction for future work.

\begin{acknowledgments}
We thank the \music team at the University of Exeter (UK) for their collaboration in developing the \music code.  Computing support for this work came from the Lawrence Livermore National Laboratory (LLNL) Institutional Computing Grand Challenge program.   This work was performed under the auspices of the U.S. Department of Energy by Lawrence Livermore National Laboratory under Contract DE-AC52-07NA27344. LLNL-JRNL-2017298.
\end{acknowledgments}


\bibliography{gofm_paper}

\end{document}